\documentclass[onecolumn]{aastex61}
\usepackage{xspace}
\usepackage{color}
\usepackage{graphicx}

\newcommand{\kms}{km s$^{-1}$\xspace}

\newcommand{\HII}{{\rm H\,{\scriptsize II}}\xspace}
\newcommand{\hii}{{\rm H\,{\scriptsize II}}\xspace}

\newcommand{\Hna}{\rm Hn$\alpha$\xspace}

\newcommand{\avHna}{$<$Hn$\alpha\!\!>$\xspace}

\shorttitle{Pilot: SHRDS}
\shortauthors{Brown, C., et al.}

\begin{document}
\title{The Southern \HII Region Discovery Survey (SHRDS): Pilot Survey}
\author{C. Brown}
\affiliation{School of Physical Sciences, Private Bag 37, University of Tasmania, Hobart, TAS, 7001, Australia}
\author{ C. Jordan}
\affiliation{ International Centre for Radio Astronomy Research, Curtin University, Perth, WA, 6845, Australia}
\author{ John M. Dickey}
\affiliation{School of Physical Sciences, Private Bag 37, University of Tasmania, Hobart, TAS, 7001, Australia}
\author{  L.D. Anderson}
\affiliation{Department of Physics and Astronomy, West Virginia University, PO Box 6315, Morgantown WV 26506, USA}
\affil{Adjunct Astronomer at the Green Bank Observatory, P.O. Box 2, Green Bank WV 24944, USA}
\affil{Center for Gravitational Waves and Cosmology, West Virginia University, Chestnut Ridge Research Building, Morgantown, WV 26505, USA}
\author{ W. P. Armentrout}
\affiliation{Department of Physics and Astronomy, West Virginia University, PO Box 6315, Morgantown WV 26506, USA}
\altaffiliation{Center for Gravitational Waves and Cosmology, West Virginia University, Chestnut Ridge Research Building, Morgantown, WV 26505, USA}
\author{ Dana S. Balser}
\affiliation{National Radio Astronomy Observatory, 520 Edgemont Road, Charlottesville, VA, 22904, USA}
\author{ T.M. Bania}
\affiliation{ Institute for Astrophysical Research, Department of Astronomy, Boston University, 725 Commonwealth Avenue, Boston, MA, 02215, USA}
\author{ J.R. Dawson}
\affiliation{ Department of Physics and Astronomy and MQ Research Centre in Astronomy, Astrophysics and Astrophotonics, Macquarie University, NSW, 2109, Australia}
\affiliation{ CSIRO Astronomy and Space Science, PO Box 76, Epping, NSW, 1710, Australia}
\author{ N.M. Mc Clure-Griffiths}
\affiliation{Research School of Astronomy \& Astrophysics, The Australian National University, Canberra ACT 2611, Australia}
\author{ Trey V. Wenger}
\affiliation{Department of Astronomy, University of Virginia, PO Box 400325, Charlottesville VA 22904-4325, USA}
\affiliation{National Radio Astronomy Observatory, 520 Edgemont Road, Charlottesville, VA, 22904, USA}

\begin{abstract}
The Southern \HII Region Discovery Survey (SHRDS) is a survey of the third and
fourth quadrants of the Galactic plane that will detect radio recombination line and continuum emission at cm-wavelengths from several hundred \hii region candidates using the
Australia Telescope Compact Array.
The targets for this survey come from the WISE Catalog of Galactic HII Regions, and were identified based on mid-infrared and radio continuum emission.
In this pilot project, two different configurations of the Compact Array Broad
Band receiver and spectrometer system were used for short test observations.
The pilot surveys detected radio recombination line emission from 36 of 53 \hii region candidates, as well as seven known \hii regions that were included for calibration.
These 36 recombination line detections confirm that the candidates are true \hii regions, and allow us to estimate their distances.
\end{abstract}


\section{Introduction}
\HII regions are zones of ionized gas surrounding young ($\sim10$ Myr old), massive stars.  They are some of the brightest objects in the Galaxy at infrared and radio wavelengths, and so they can be detected across the entire Galactic disk \citep{Anderson11}.  \hii regions are the archetypical tracers of Galactic spiral structure and their chemical abundances provide unique and important probes of billions of years of Galactic chemical evolution \citep{Shaver83}.  They are the main tracer of ionizing photons in the Galaxy, and can be used to compute global star formation rates.  Unlike most tracers of high-mass star formation (e.g. far-infrared clumps), \HII regions unambiguously locate sites where massive stars have recently formed. An unbiased Galaxy-wide sample of \HII regions is required to understand the global properties of the Milky Way, and to compare its star formation rate to those of external galaxies.\\ 

The \HII Region Discovery Survey (HRDS) is a collection of radio recombination line (RRL) and continuum emission surveys between 4 and 11 GHz, designed to detect all \HII regions ionised by single or multiple O-stars across the entire Galactic disk.  All HRDS component surveys are targeted towards \HII region candidates, selected to have spatially coincident $\sim25$ $\micron$ mid-infrared and $\sim20$ cm radio continuum emission, surrounded by $\sim10$ $\micron$ emission---a basic morphology shared by all Galactic \HII regions \citep{Anderson14}.  But these criteria are not sufficiently robust to exclude all other kinds of radio and infrared sources; to confirm that they are \HII regions and to measure their radial velocities, it is necessary to detect RRL emission from each candidate. \\

The primary instrument used for the HRDS is the Green Bank Telescope \citep[GBT,][]{GBTHRDS}.  Spanning $-16\arcdeg<\ell<67\arcdeg$ and $|b|<1\arcdeg$, the GBT HRDS detected 602 discrete recombination line components from 448 pointings. This more than doubled the number of known \HII regions in this part of the Galaxy. Continuing the GBT HRDS, the Arecibo HRDS \citep{Bania12} discovered 37 previously unknown \HII regions in the area $31\arcdeg<\ell<66\arcdeg$, $|b|<1\arcdeg$. Recently, the GBT HRDS has been extended by \citet{Anderson15}, resulting in a further 302 \HII region discoveries.  
Together, these three northern HRDS surveys achieve a detection rate $>90$\%, resulting in the discovery of nearly 800 \HII regions---including the most distant Galactic \HII regions known.\\

The Southern \HII Region Discovery Survey (SHRDS) is a multi-year project using the Australia Telescope Compact Array (ATCA) to complement the GBT and Arecibo HRDS by extending the survey area into the southern sky ($\delta<-45\arcdeg$). This area includes the Southern end of the Galactic Bar, the Near and Far 3kpc Arms, the Norma/Cygnus Arm, the Scutum/Crux Arm, the Sagitttarius/Carina Arm, 
and outside the solar circle, the Perseus Arm, and the Outer Arm.
Our lists of confirmed \HII regions are seriously incomplete in the third and fourth Galactic quadrants, where no large-scale \HII region RRL survey has been done in nearly three decades \citep[since the work of ][]{CH87}. Currently, three candidate \HII regions exist for every confirmed \HII region between $225\arcdeg<\ell<340\arcdeg$, i.e. outside the declination range observable with the GBT.\\

We make use of the ATCA's compact array configurations, wide band C/X receiver and Compact Array Broadband Backend \citep[CABB, ][]{CABB} to observe up to 25 $\alpha$ RRL transitions simultaneously. The transitions and polarisations can be averaged together in order to produce a single average \avHna spectrum, providing roughly a factor of five improvement in sensitivity compared with a conventional single line observation. With a moderate integration time per candidate, the ATCA can improve the detection threshold of a RRL survey by nearly an order of magnitude compared to the \citet{CH87} Parkes catalog.\\

The analysis of the pilot survey data was done entirely on the \emph{uv} plane.  The sparse 
\emph{uv} coverage for each candidate is not sufficient to make images with good dynamic range.
The full SHRDS survey will collect data with multiple telescope configurations and many 
hour-angle scans on each source, so that maps of the continuum and spectral line emission 
with good resolution and fidelity can be obtained. \\

This paper presents the results of two SHRDS pilot observing sessions, in 2013 and 2014, and introduces the telescope and receiver configuration used for the SHRDS.  The Pilot Survey source selection, observation, and data reduction strategies are discussed in Sections \ref{sec:SourceSelection},  \ref{sec:Observations} and \ref{sec:DataReduction}. The results of the pilot observations are presented on 
Table 3.



\section{Source Selection}\label{sec:SourceSelection}

The HRDS and SHRDS are not blind surveys complete over defined areas, instead the survey observations are targeted towards \HII region candidates. Candidate selection is based on infrared and radio continuum morphology, as discussed by \citet{Anderson14}.  
In the third and fourth Galactic quadrants, the mid-infrared data comes from the WISE catalog
of Galactic \hii regions, which contains roughly 2000 radio-loud candidates.
In addition to the WISE \citep{WISE} data at 12 and 22 $\mu$m wavelength
we also consider Spitzer GLIMPSE at 8 $\mu$m
\citep{Benjamin2003, Churchwell2009} and Spitzer MIPSGAL at 24 $\mu$m \citep{Carey2009}. The radio continuum data comes primarily
from the SUMSS survey \citep{SUMSS}, with reference also to the MAGPIS, NVSS and SGPS 
surveys \citep[][respectively]{MAGPIS, NVSS, SGPS}.  To predict the flux density at 
$\lambda$ 6 cm we assumed an optically thin spectral index of $\alpha = -0.12$. 
The list of targets for the pilot observations is given on Table 1.

\startlongtable
\begin{deluxetable*}{lccccc}
\tablecolumns{6}
\tablewidth{0pt}
\tablecaption{Pilot SHRDS Source Candidates.  The region name, Epoch
  of observation, number of $uv$ cuts, and total observation time are
  listed for each source.  The final column
  indicates that there are known velocities (e.g. stellar or
  molecular) probably associated with the region.}
\label{table:ObservedSources}
\tablehead{
\colhead{Source} & \colhead{Obs.} & \colhead{\#$uv$} & \colhead{Int.} &  & \colhead{Known} \\
\colhead{Name} & \colhead{Epoch} & \colhead{cuts} & \colhead{(min)} & \colhead{IRAS source}  & \colhead{Velocities}\\
}
\startdata
\cutinhead{\HII Region Candidates}
G$      214.250-02.461  $&      2       &       6       &       25      &       \object{IRAS 06425-0214}        &                \nodata        \\
G$      217.497-00.008  $&      2       &       6       &       25      &        \object{IRAS 06571-0359}       &               \nodata         \\
G$      217.640-00.057  $&      2       &       6       &       25      &        \object{IRAS 06573-0408}       &               \nodata         \\
G$      222.096-01.981  $&      2       &       6       &       25      &        \object{IRAS 06587-0859}       &               Y       \\
G$      222.159-02.163  $&      2       &       6       &       25      &        \nodata        &        Y       \\
G$      230.354-00.597  $&      2       &       7       &       28      &        \object{IRAS 07195-1538}       &               \nodata         \\
G$      234.267-01.496  $&      2       &       7       &       28      &        \object{IRAS 07240-1930}       &       \nodata \\ 
G$      234.673-00.243  $&      2       &       7       &       28      &        \object{IRAS 07295-1915}       &        Y       \\
G$      234.762-00.277  $&      2       &       7       &       28      &        \object{IRAS 07296-1921}       &        Y       \\
G$      235.696-01.243  $&      2       &       7       &       28      &        \object{IRAS 07279-2038}       &        Y       \\
G$      237.232-01.066  $&      2       &       7       &       28      &        \object{IRAS 07318-2153}       &        Y       \\
G$      237.257-01.281  $&      2       &       7       &       28      &       \object{IRAS 07310-2201}        &        Y       \\
G$      239.332-02.738  $&      2       &       7       &       28      &        \object{IRAS 07299-2432}       &        Y       \\
G$      290.012-00.867  $&      2       &       7       &       25      &        \object{IRAS 10595-6041}       &        \nodata         \\
G$      290.385-01.042  $&      2       &       6       &       25      &        \nodata        &       \nodata \\
G$      290.674-00.133  $&      2       &       6       &       25      &        \object{IRAS 11069-6016}       &        \nodata         \\
G$      291.596-00.239  $&      2       &       6       &       25      &        \object{IRAS 11137-0239}       &        \nodata         \\
G$      292.722+00.157  $&      2       &       7       &       25      &        \object{IRAS 11233-6043}       &        \nodata         \\
G$      292.889-00.831  $&      2       &       7       &       25      &        \object{IRAS 11220-6142}       &        \nodata         \\
G$      293.483-00.903  $&      2       &       5       &       25      &        \object{IRAS 11265-6158}       &        Y       \\
G$      293.936-00.873  $&      2       &       6       &       25      &        \nodata        &       Y       \\
G$      293.994-00.934  $&      2       &       6       &       25      &       \nodata &       Y       \\
G$      294.656-00.438  $&      2       &       13      &       52      &        \nodata        &       \nodata         \\
G$      294.988-00.538  $&      2       &       12      &       48      &        \object{IRAS 11396-6202}       &       Y       \\
G$      295.275-00.255  $&      2       &       7       &       25      &        \object{IRAS 11427-6151}       &        \nodata         \\
G$      297.248-00.754  $&      2       &       7       &       28      &        \object{IRAS 11583-6247}       &        \nodata        \\
G$      297.626-00.906  $&      2       &       7       &       28      &        \object{IRAS 12013-6300}       &        \nodata         \\
G$      298.473+00.104  $&      2       &       7       &       28      &       \nodata &       \nodata \\
G$      298.669+00.064  $&      2       &       7       &       28      &        \object{IRAS 12117-6213}       &        \nodata         \\
G$      299.505+00.025  $&      2       &       7       &       28      &        \nodata        &        \nodata         \\
G$      300.519-00.409  $&      2       &       7       &       28      &        \object{IRAS 12271-6253}       &        \nodata         \\
G$      300.972+00.994  $&      2       &       7       &       29      &        \object{IRAS 12321-6132}       &        \nodata         \\
G$      300.983+01.117  $&      2       &       7       &       28      &        \object{IRAS 12320-6122}       &        Y       \\
G$      313.671-00.105  $&      1       &       5       &       51      &        \object{IRAS 14183-6050}       &        Y       \\
G$      314.219+00.344  $&      1       &       4       &       14      &        \nodata        &        Y       \\
G$      316.516-00.600  $&      1       &       4       &       40      &        \object{IRAS 14412-6013}       &        Y       \\
G$      317.861+00.160  $&      1       &       3       &       10      &        \object{IRAS 14482-5857}       &        Y       \\
G$      318.248+00.151  $&      1       &       3       &       15      &        \nodata        &       \nodata \\
G$      319.229+00.225  $&      1       &       3       &       31      &        \nodata        &        \nodata \\
G$      323.449+00.095  $&      1       &       4       &       22      &        \object{IRAS 15246-5612}       &        Y       \\
G$      323.743-00.249  $&      1       &       5       &       31      &        \object{IRAS 15278-5620}       &        Y       \\
G$      323.806+00.020  $&      1       &       4       &       16      &        \object{IRAS 15270-5604}       &        \nodata \\
G$      323.936-00.037  $&      1       &       3       &       19      &        \nodata        &        \nodata \\
G$      324.662-00.331  $&      1       &       3       &       10      &        \object{IRAS 15335-5552}       &        \nodata        \\
G$      325.108+00.054  $&      1       &       3       &       22      &        \object{IRAS 15347-5518}       &        \nodata \\
G$      325.354+00.035  $&      1       &       4       &       16      &        \object{IRAS 15392-5545}      &       \nodata \\
G$      326.721+00.773  $&      1       &       3       &       10      &        \object{IRAS 15404-5345}       &        Y       \\
G$      326.890-00.277  $&      1       &       3       &       10      &        \object{IRAS 15457-5429}       &        Y       \\
G$      326.916-01.100  $&      1       &       3       &       10      &        \object{IRAS 15495-5505}       &        Y       \\
G$      327.401+00.484  $&      1       &       3       &       10      &        \object{IRAS 15454-5335}       &        Y       \\
G$      327.555-00.829  $&      1       &       3       &       10      &       \nodata &       Y       \\
G$      327.714+00.576  $&      1       &       3       &       31      & \nodata        &       \nodata \\
G$      327.763-00.163  $&      1       &       3       &       17      &       \nodata &       \nodata \\
\cutinhead{Known \HII Regions}
G$      213.833+00.618  $&      2       &       6       &       25      &        \object{IRAS 06527-0027}       &        Y       \\
G$      290.323-02.984  $&      2       &       7       &       25      &        \object{IRAS 10545-6244}       &        Y       \\
G$      295.748-00.207  $&      2       &       7       &       28      &        \object{IRAS 11467-6155}       &        Y       \\
G$      315.312-00.273  $&      1       &       4       &       14      &        \nodata        &        Y       \\
G$      313.790+00.706  $&      1       &       5       &       14      &        \object{IRAS 14170-6002}       &        Y       \\
G$      323.464-00.079  $&      1       &       4       &       40      &        \object{IRAS 15254-5621}       &        Y\\
G$      327.313-00.536  $&      1       &       3       &       11      &        \object{IRAS 15492-5426}       &        Y      \\
\enddata
\end{deluxetable*}

\subsection {Pilot Survey Source Selection}
The SHRDS pilot observations were done in two sessions.  Epoch I, observed June 30, 2013, focused on candidates that were expected to show bright RRL detections, which they did.  Epoch II, observed June 26 and 27, 2014, used a list of candidates with expected flux densities typical of the SHRDS catalog as a whole.  The two epochs also used different longitude ranges in order
to generate samples of \hii regions with different Galactic radii. 

\subsubsection{Epoch I}
For the first round of observations, we observed \HII region candidates in the range $312\arcdeg<\ell<328\arcdeg$.  This section of the fourth Galactic quadrant
provides lines of sight near the tangents to the Norma-Cygus and Scutum-Crux arms in the inner Galaxy, and roughly perpendicular to the Scutum-Centaurus and Sagittarius Arm tangents in the outer Galaxy.\\

We selected twenty \HII region candidates \citep[from ][]{Anderson14} for observations in Epoch I. In addition, four known \HII regions: \object[GAL 315.31$-$00.27]{G315.312$-$00.272} and \object[GAL 327.30$-$00.60]{G327.300$-$00.548} \citep[from ][]{CH87}, and G313.790+00.706 and G323.464$-$00.079 \citep[from ][]{Misanovic02}, were included in the observation schedule.
Thus Epoch I included a total of 24 targets (Table 1).
RRLs from all candidates and known regions were detected.

\subsubsection{Epoch II}
The observations in Epoch II covered longitude range $213\arcdeg<\ell<301\arcdeg$.
This area selects mostly 
\HII regions with Galactocentric radii outside the solar circle.
This part of the disk has been little studied in previous Galactic RRL surveys.  Between $213\arcdeg<\ell<301\arcdeg$ the ratio of known:candidate \HII regions is 2:5, compared with 1:1 
in the corresponding longitude range in the first and second quadrants ($59\arcdeg<\ell<147\arcdeg$) for candidates selected according to the same criteria.
Overall the detection rate for the \hii region candidates 
observed in epoch II was 48\% (16 out of 33). Of the 13 candidates in the third quadrant, only one was detected, G230.354$-$00.597, plus the known source G213.833+00.618.  In both cases the lines are only just above the 3$\sigma$ threshold.\\

Epoch II candidates were selected to have WISE radii between $60\arcsec$ and 150$\arcsec$ and no known radial velocity information
as tabulated by \citet{Anderson14}. We selected 33 \HII region candidates between $213\arcdeg<\ell<240\arcdeg$ and $290\arcdeg<\ell<301\arcdeg$ that fulfilled these criteria.  Therefore Epoch II observed a more representative sample from the catalog of \citet{Anderson14} to determine detection statistics for the SHRDS full survey.\\

As in Epoch I, a few known \HII regions were added to the observation schedule: G213.883+00.618, G290.323$-$02.984 and G295.748$-$00.207.  Of these three,
G295.748$-$00.207 was strongly detected, and G213.883+00.618 and G290.323$-$02.984 were detected at the 3.5$\sigma$ and 5.6$\sigma$ levels, respectively.

\section{Observations}\label{sec:Observations}

\begin{figure}
\centering
\includegraphics[width=4in]{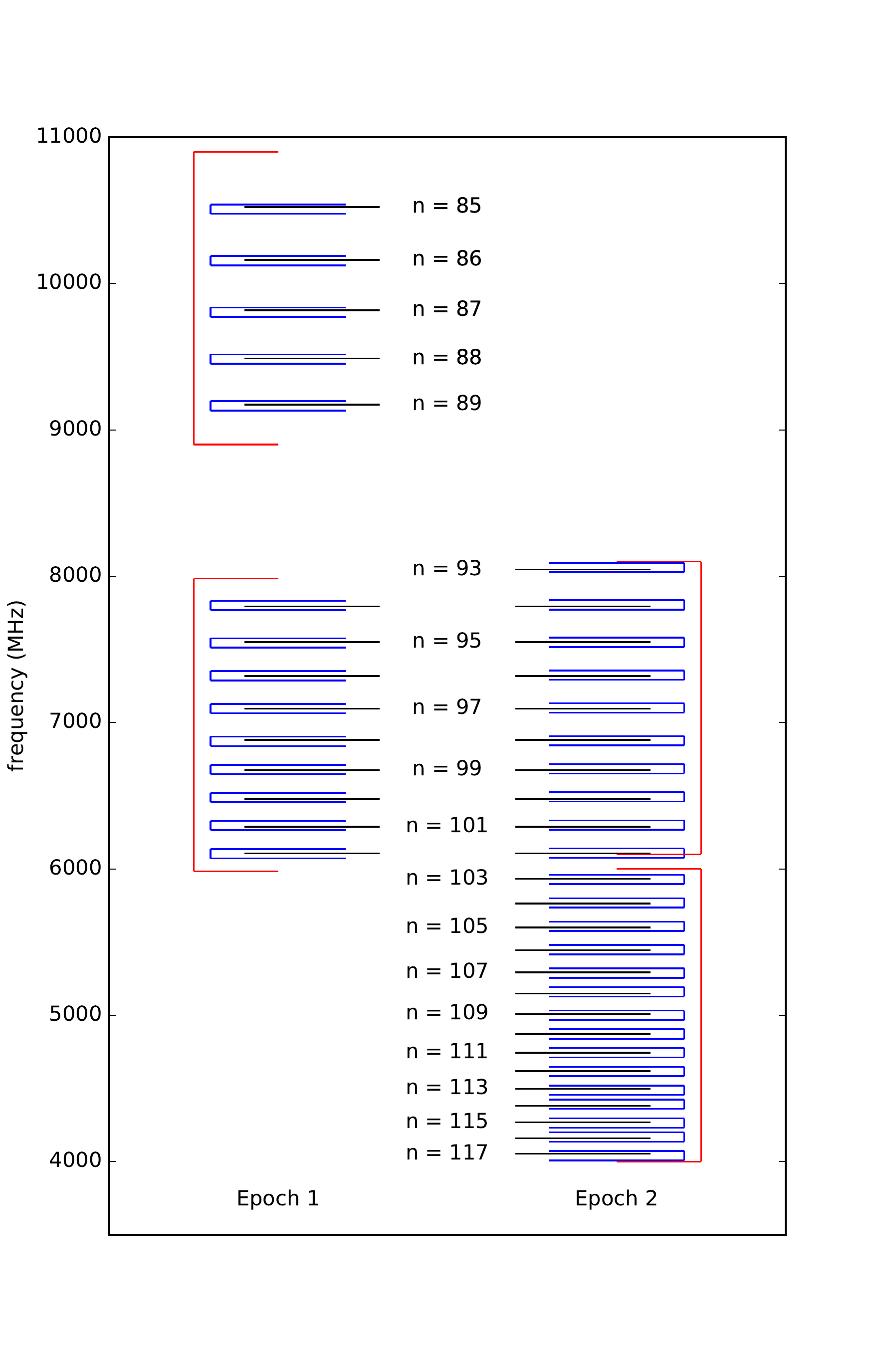}
\caption{A schematic of CABB frequencies for Epoch I (left) and Epoch II (right).  The two 2 GHz CABB bands are shown in red, the individual zoom bands are shown in blue.  The frequencies of the \Hna recombination lines are shown as black horizontal lines, with the level n indicated. \label{CABBschematic}}
\end{figure}

\startlongtable
\begin{deluxetable}{cccc}
\tablecolumns{4}
\tablewidth{0pt}
\tablecaption{Recombination lines observed. Hn$\alpha$ transitions and
  rest frequencies are given, with the center frequency and velocity
  resolution of each zoom band.  The center frequencies of the 2-GHz
  bands ($\nu_1$ and $\nu_2$) are also listed. See Figure
  \ref{CABBschematic}. \label{Table:ZoomSetup} }
\tablehead{
\colhead{Hn$\alpha$} & \colhead{$\nu$ rest} & \colhead{$\nu$ central} & \colhead{$\Delta V$} \\
\colhead{$n=$} & \colhead{MHz} & \colhead{MHz} & \colhead{\kms}
}
\startdata
\cutinhead{Epoch I, $\nu_1 = 6984$, $\nu_2 = 9900$ MHz}
85      &       10522.04        &       10508   &       0.9     \\
86      &       10161.30        &       10156   &       0.9     \\
87      &       9816.864        &       9804    &       1.0     \\
88      &       9487.821        &       9484    &       1.0     \\
89      &       9173.321        &       9164    &       1.0     \\
94      &       7792.871        &       7800    &       1.2     \\
95      &       7550.614        &       7544    &       1.2     \\
96      &       7318.296        &       7320    &       1.3     \\
97      &       7095.411        &       7096    &       1.3     \\
98      &       6881.486        &       6872    &       1.4     \\
99      &       6676.076        &       6680    &       1.4     \\
100     &       6478.760        &       6488    &       1.4     \\
101     &       6289.144        &       6296    &       1.5     \\
102     &       6106.855        &       6104    &       1.5     \\
\cutinhead{Epoch II, $\nu_1 = 5000 $, $\nu_2 = 7100$ MHz}
93      &       8045.603        &       8060    &       1.2     \\
94      &       7792.871        &       7804    &       1.2     \\
95      &       7550.614        &       7548    &       1.2     \\
96      &       7318.296        &       7324    &       1.3     \\
97      &       7095.411        &       7100    &       1.3     \\
98      &       6881.486        &       6876    &       1.4     \\
99      &       6676.076        &       6684    &       1.4     \\
100     &       6478.760        &       6492    &       1.4     \\
101     &       6289.144        &       6300    &       1.5     \\
102     &       6106.855        &       6108    &       1.5     \\
103     &       5931.544        &       5928    &       1.6     \\
104     &       5762.880        &       5768    &       1.6     \\
105     &       5600.550        &       5608    &       1.7     \\
106     &       5444.260        &       5448    &       1.7     \\
107     &       5293.732        &       5288    &       1.8     \\
108     &       5148.703        &       5160    &       1.8     \\
109     &       5008.923        &       5000    &       1.9     \\
110     &       4874.157        &       4872    &       1.9     \\
111     &       4744.776        &       4744    &       2.0     \\
112     &       4618.789        &       4616    &       2.0     \\
113     &       4497.776        &       4488    &       2.1     \\
114     &       4380.954        &       4392    &       2.1     \\
115     &       4268.142        &       4264    &       2.2     \\
116     &       4159.171        &       4168    &       2.2     \\
117     &       4053.878        &       4040    &       2.3     \\
\enddata
\end{deluxetable}

All pilot SHRDS observations used the ATCA in the five antenna H75 array
configuration, giving a nominal maximum baseline of 75m
and a beam size of FWHM$\simeq$
65\arcsec \ at 7.8 GHz depending on the declination and hour
angles of the observations.  As an interferometer survey, the SHRDS 
cannot detect emission spread smoothly over much larger angles than the
shortest projected baseline, which can be as short as the dish diameter, 22 m.
This largest angular scale is roughly equal to the primary beam size,
FWHM$= 6 \arcmin$ at 7.8 GHz.  Although the resolution is very coarse, the
H75 configuration gives the best brightness temperature sensitivity, which
is the critical parameter for detecting weak spectral line emission from
extended sources like most \hii regions.
Surveys of compact and ultra-compact \hii regions, like the
CORNISH and SCORPIO surveys \citep[]{Hoare12, Purcell13,
Umana15} use very different telescope configurations.\\

The ATCA's Compact Array Broadband Backend \citep[CABB, ][]{CABB} and C/X
upgrade allow for two 2-GHz spectral windows to be placed anywhere between
4.0 and 10.8 GHz.  The 64M-32k observing mode used here provides for each of
these two windows a coarse resolution spectrum of  
32 x 64-MHz channels and up to 16 fine resolution ``zoom'' bands
of 2048 channels across 64 MHz, placed within each broadband 2 GHz window. 
The zoom bands provide very high spectral resolution, with channel separation
32 KHz = 1.2 \kms at 7.8 GHz, and velocity range of
nearly 2500 \kms each.  Thus it is not
necessary to center the line rest frequency in each zoom band.  The center
frequencies of the zoom bands are constrained to have frequency
separations equal to integral multiples of half the zoom band width.
In practice, selection of zoom band center frequencies is facilitated by 
the CABB scheduler, part of the ATCA observation scheduling tool
(\url{https://www.narrabri.atnf.csiro.au/observing/sched/cabb/}).
After calibration and Doppler correction, the zoom bands can be aligned in LSR velocity and
resampled to the same channel spacing, and then averaged in order to improve the signal to noise ratio for weak and marginal detections.\\

There are 33 \Hna RRL transitions within the frequency range of the 4 cm
receiver. However, the H86$\alpha$ line is spectrally compromised by
higher order RRL transitions \citep{Balser06}, and the
H90$\alpha$ transition can be affected by a trapped mode in the
ATCA's 6/3cm horn. This leaves 31 individual \Hna transitions between
4.0 and 10.8 GHz.  The placement of the two 2-GHz CABB bands, and hence the selection of which RRL transitions to observe,
is complicated by interference at many frequencies, by variations
in the system temperature of the receiver with frequency, and by the
natural decrease of the line-to-continuum ratio with increasing quantum level.  
The two epochs of pilot observations explored two among many possible choices of CABB 
frequencies (see Figure \ref{CABBschematic} and Table \ref{Table:ZoomSetup}).\\

Epoch I (project C2842) was observed over twelve hours on July 1, 2013. The frequency placement
of the various bands for Epoch I was chosen to emulate the GBT HRDS
observations. 
Centering the 2 GHz CABB bands at 7000 MHz and 9900 MHz allows thirteen \Hna RRLs to be observed simultaneously.  This frequency range includes three of the RRLs observed by the GBT HRDS, and an additional ten \Hna transitions (see Figure \ref{CABBschematic}).
A single phase calibrator, \object{PKS B1421$-$490}, was used
for all Epoch I observations.
Both Epochs used \object{PKS 1934$-$63} as a flux calibrator, and phase 
calibration was done every $\sim20$ minutes.  Each source was
observed at several hour angles to improve the sampling on the \emph{uv} plane.\\

Epoch II (project C2963) was observed over two twelve hour blocks on July 26 and 27, 2014.  The centers of the two broadband IFs were moved to 5000 and 7100 MHz in order to increase the number of simultaneously observed \Hna transitions from 13 to 25.  
In addition to PKS1934$-$638, secondary bandpass calibrators (\object{PKS0823$-$500} and \object{PKS0537$-$441}) were also observed in Epoch II.
As the Epoch II targets fell into two longitude groups ($\ell\approx220\arcdeg$ and $\ell\approx295\arcdeg$), two phase calibration sources were chosen for each day: \object[PKS0723-008]{PKS0723$-$008} and \object[PMS J1131-5818]{PMS J1131$-$58} for July 26, and \object[PKS 0727-11]{PKS0727$-$115} and \object[PKS 1148-671]{PKS1148$-$671} for July 27.\\

The lower frequency band centers selected in Epoch II did not give good results, even though
they cover more RRL $\alpha$ transitions.  There is more artificial interference
at frequencies below 5 GHz, and several of the zoom bands had to be discarded entirely.
The conclusion for the subsequent SHRDS survey is that the best frequency placement is
a compromise between the two shown on Figure \ref{CABBschematic}.  For the full survey
we chose to center our broad bands at 5.505 and 8.540 GHz.  In this configuration the SHRDS can observe 18 RRL $\alpha$ transitions simultaneously.

\section{Data Processing and Analysis}\label{sec:DataReduction}

Bandpass calibration, flux density calibration and flagging were carried out with standard \textsc{miriad} reduction techniques \citep{MIRIAD}.
After calibration, the zoom bands can be aligned in velocity and
resampled to the same channel spacing, and then averaged in order to improve the signal to noise ratio.
For Epoch I we set the common channel spacing to $\Delta$V = 2.5 \kms, and in Epoch II we
use 2.3 \kms (Table \ref{Table:ZoomSetup}).  The resampling was done with the \textsc{miriad} task \textsc{uvaver}, that uses Fourier extension to change the channel step size.
We average \emph{uv} spectra weighting by the continuum flux density of each baseline and each band.  Longer
baselines generally give weaker continuum since most \hii regions are partially resolved even with the H75 array.
Any individual transitions that were polluted by RFI or had bad baseline ripples due to calibration problems were not
included in the final average spectra.\\

Working with \emph{uv} plane data can blend together spectra from multiple objects
within the primary beam.  To separate individual sources or source conponents in 
a crowded field requires imaging.  This analysis will follow when the full 
SHRDS survey data are available.\\ 

 \begin{figure*}
        \centering
        \includegraphics[width=3in]{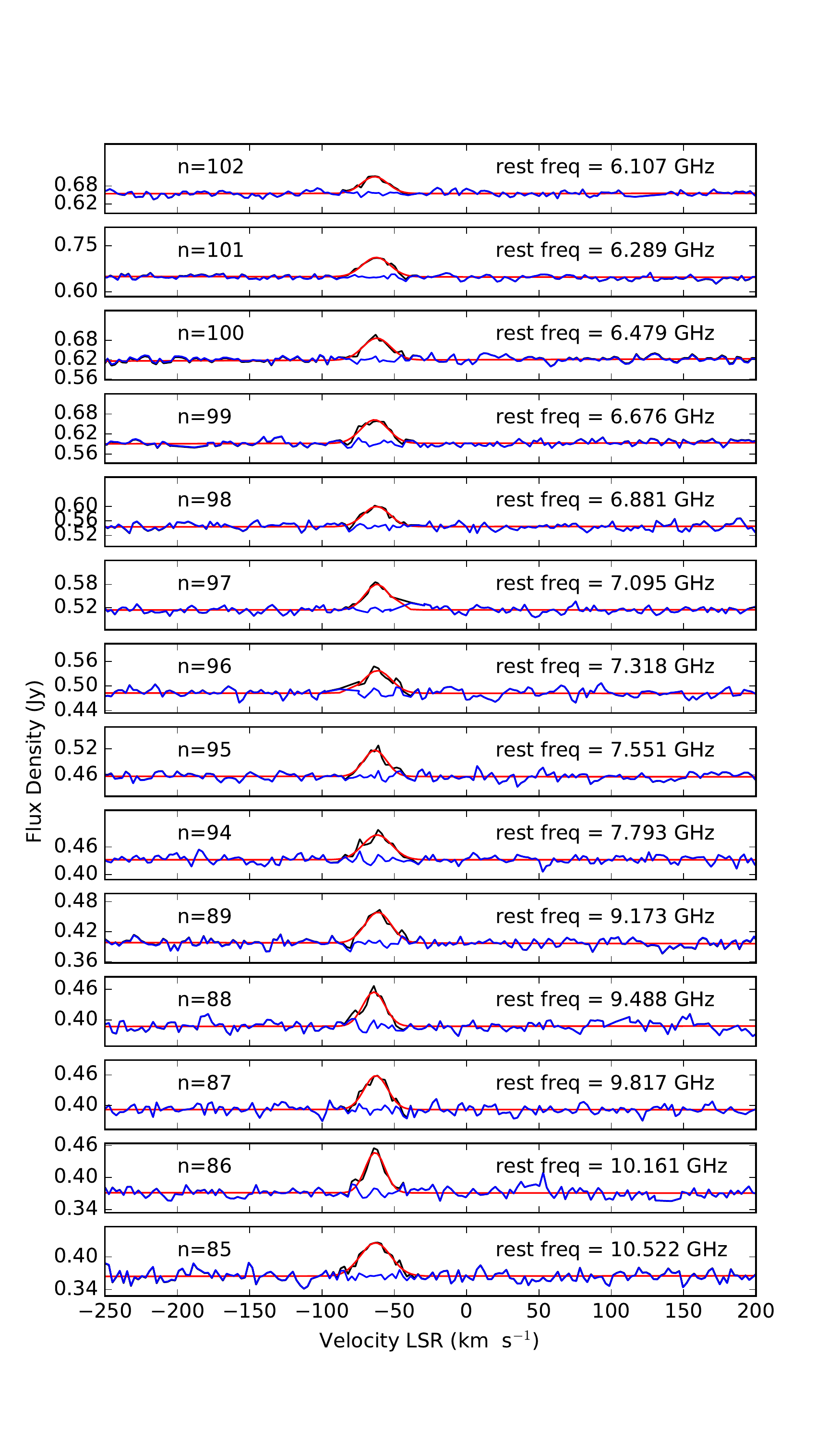}
        \includegraphics[width=3in]{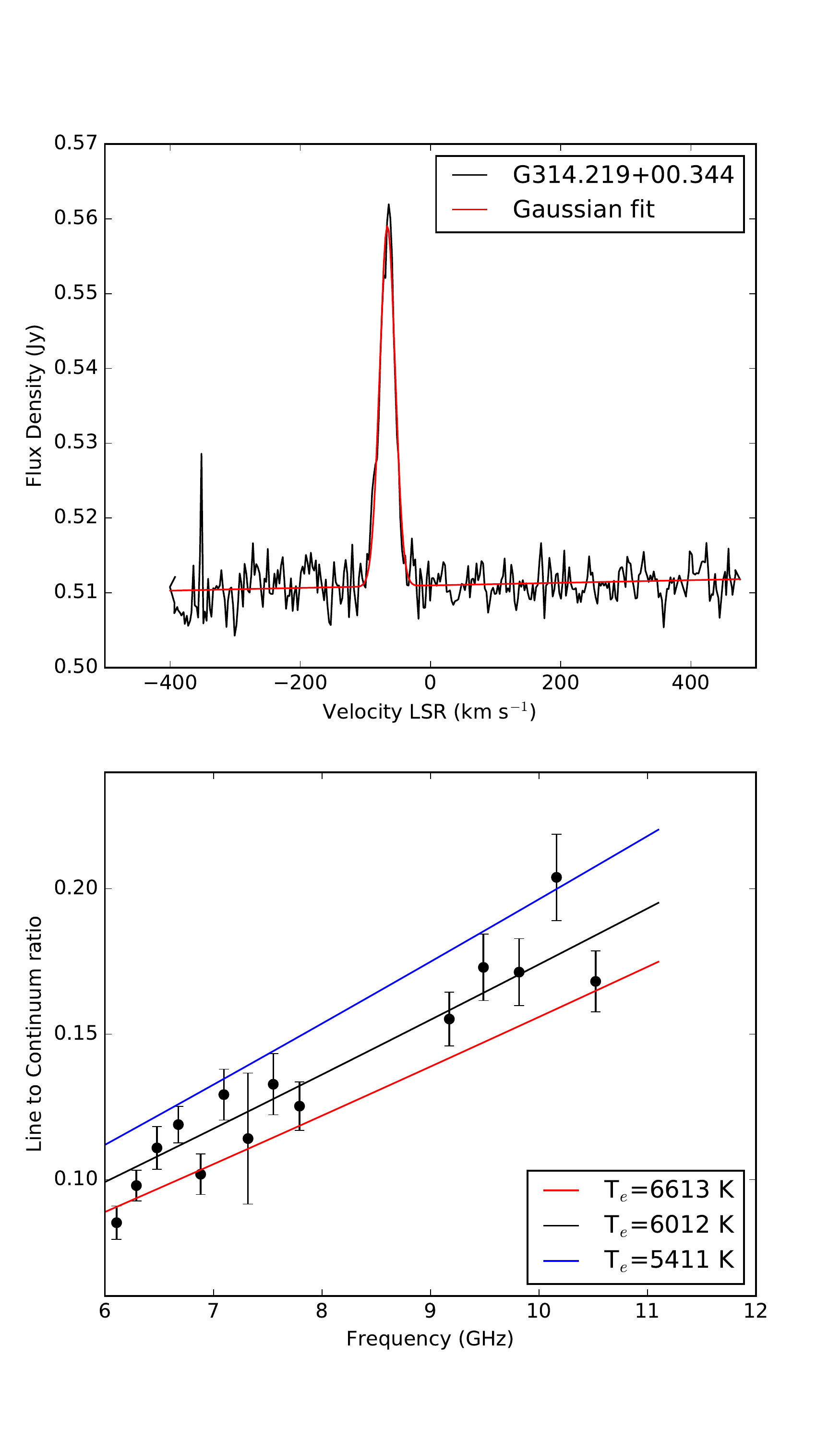}
        \caption{Spectra from a bright but previously unknown source.  
This figure shows the spectra for G314.219+00.344, which is a moderate to strong
source, not previously detected in RRL emission. Individual RRL transitions are shown on the left panel, the overall average with Gaussian fit is shown on the upper right. The lower-right panel shows the line-to-continuum ratio for each RRL transition vs.
frequency.  Fitting these data with the assumption that both the line and continuum
emission are optically thin gives an electron temperature T$_e \ \simeq$ 6000 K
(black line).  The red and blue lines correspond to increase and
decrease by 10\% of the best fit T$_e$, as indicated.} \label{SourceFig1}
 \end{figure*}

 \begin{figure*}
        \centering
        \includegraphics[width=3in]{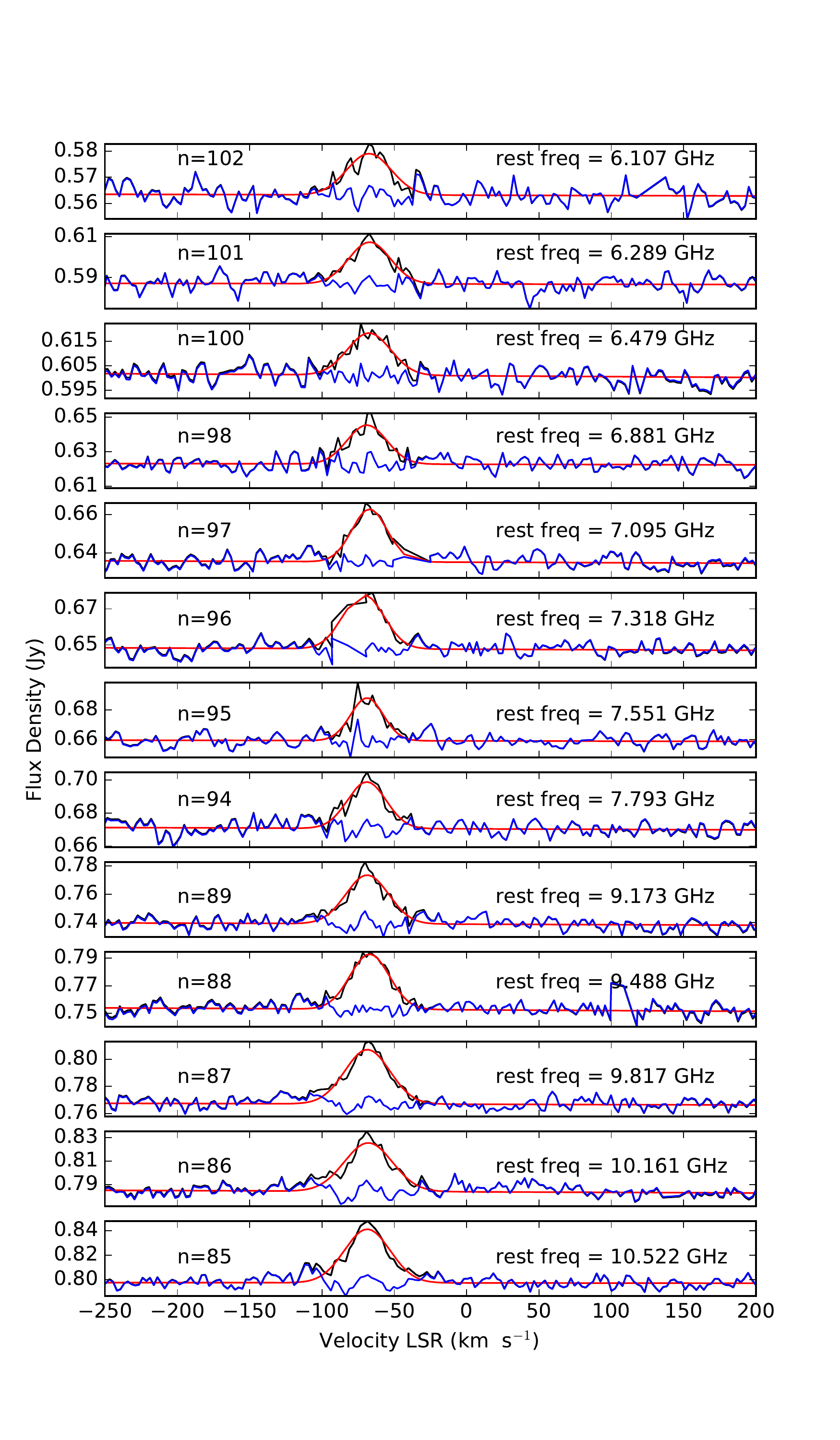}
        \includegraphics[width=3in]{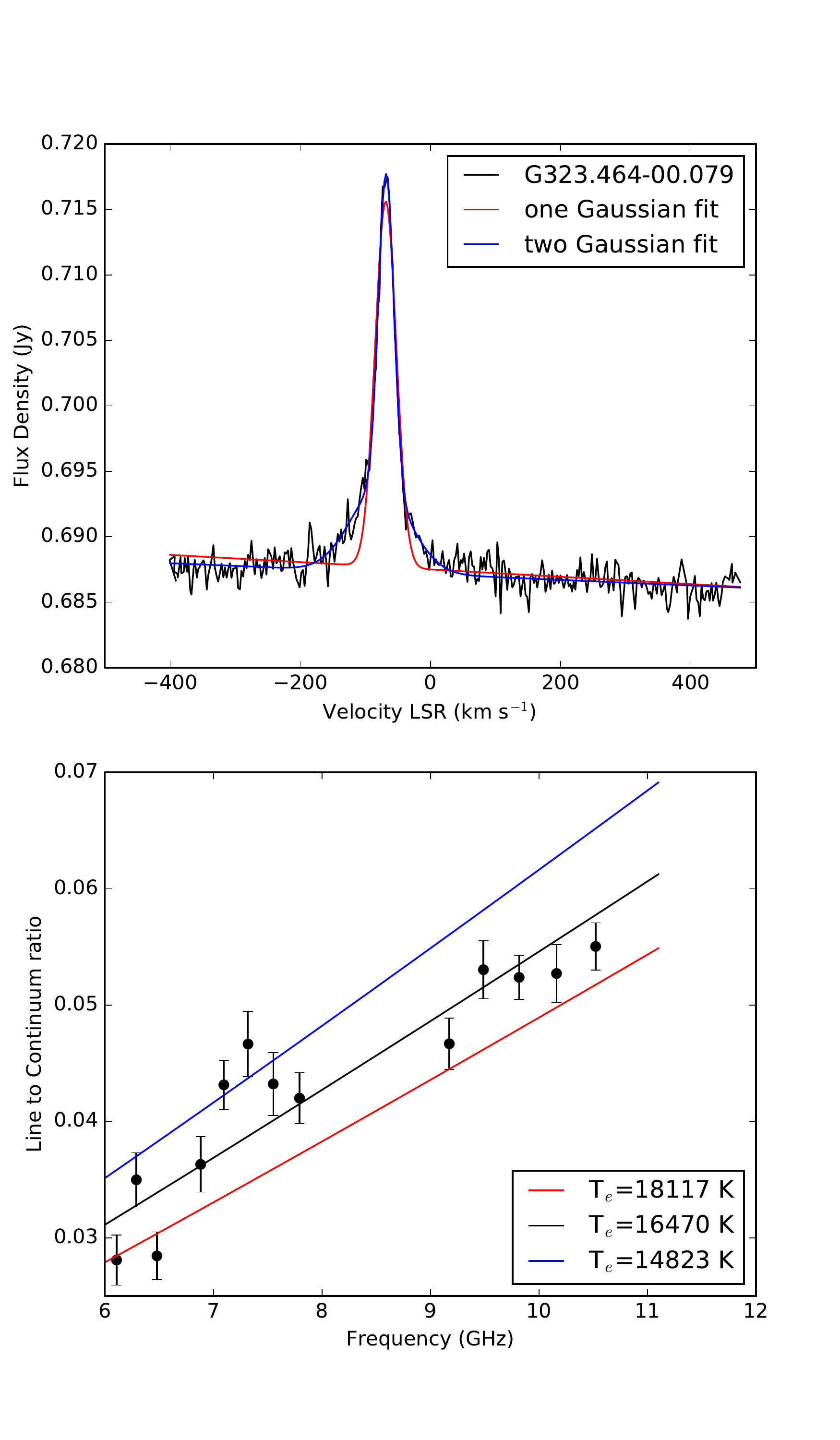}
        \caption{Spectra for a bright source that shows multiple linewidths. 
This figure shows the spectra for G323.464$-$00.079, which is a strong
source, previously detected by \citet{Murphy10}, with panel layout
the same as in Figure \ref{SourceFig1}. 
This hypercompact \hii region has a broad line profile that is not
adequately fitted by a single Gaussian.  This may indicate that there are two
ionized regions with very different kinetic temperatures in the same \hii region
complex. 
{\bf The results of the two-Gaussian fit are given on Table \ref{Table:ResultsTable}, and
indicated by ``S2'' in column 2.}
} \label{SourceFig2}
 \end{figure*}

A Gaussian fit was made to the line profile in the average spectrum for each candidate.  
For lines with SNR$>$3,
the Gaussian parameters are given on
Table \ref{Table:ResultsTable}, and illustrated on Figures
\ref{SourceFig1} and \ref{SourceFig2}.
Table \ref{Table:ResultsTable} gives for each detected source, the source name in column 1). 
{\bf Column 2 has an
S (for ``Source Average'') if the detection was made in the average of all RRL transitions or
the Hn$\alpha$ number for sources bright enough to be 
detected in individual transitions. The following columns give
}
the line center velocity (VLSR), line width (FWHM), continuum flux density ($S_C$), spectral rms noise,
peak line flux density ($S_L$), electron temperature estimated from the line-to-continuum ratios ($T_e$), and the signal-to-noise ratio, SNR.
The SNR is computed as \citep[e.g.][]{LenzAyers92}:
\begin{equation} SNR \ = \ 0.7 \ \left(\frac{S_L}{RMS}\right) \  \sqrt{ \frac{FWHM}{\Delta V}} \end{equation}
If a line was detected with signal-to-noise ratio greater than $\sim$15 in the average
spectrum, then each individual Hn$\alpha$ transition was considered
separately, again weighting the different baselines by their continuum flux.
Gaussians were fitted to the average line profile for each transition.  The results are listed as 
separate entries under each source name in Table \ref{Table:ResultsTable}.  For these lines, the SNR is 
simply the peak line flux divided by its error.\\

For lines sufficiently strong that their peak can be determined fairly accurately for each transition
separately, it is possible to estimate the electron temperature, under the assumption that both the
spectral line and continuum emission are optically thin.   Whether or not this assumption holds depends 
on the emission measure.  Typically for diffuse \hii regions the continuum is optically thin for frequencies above 
three to five GHz, but for ultracompact and hypercompact \hii regions the continuum can be optically thick up to 
frequencies much higher than the C/X band observed here (4 to 11 GHz).  If the line and continuum are both
optically thin, and the level populations are in thermodynamic equilibrium with the electron kinetic temperature, $T_e$, then the line-to-continuum ratio, $S_L$ / $S_C$, is :
\begin{equation} \frac{S_L}{S_C} \ = \ 7 \times 10^3 \ \left( \frac{\Delta V_{FWHM}}{\mathrm{km \ s}^{-1}} \right)^{-1} \ \left( \frac{\nu}{\mathrm{GHz}} \right)^{1.1} \ \left(\frac{T_e}{{\mathrm K}} \right)^{-1.15} \ \left( 1 \ + \ \frac{N(He^+)}{N(H^+)} \right)^{-1} \label{eqTe} \end{equation}
where $\Delta V_{FWHM}$ is the line width (full width to half maximum, or 2.35 times $\sigma_v$), $\nu$ is the line rest frequency, and the ratio of column densities of He$^+$ to H$^+$ is taken as 0.09, making the final term 1.09 \citep{Quireza06}.

An interferometer telescope is particularly well suited to measurement of $T_e$, because each baseline at each frequency measures $S_L$ and $S_C$ through the same spatial filter or fringe pattern.  Although different baselines measure different continuum 
flux values, depending on the angular size and structure of the source and the projected baseline length, there is no need to determine a zero-level or overall offset to the continuum flux, as there is for single-dish surveys.  In principal, every Hn$\alpha$
transition provides a separate measurement of $T_e$ by equation \ref{eqTe} (column 8 on table \ref{Table:ResultsTable}).  A better estimate is given by fitting the line-to-continuum ratios for all lines to a single $T_e$ value.  This is shown for the two sample sources
on the lower right panels of Figures \ref{SourceFig1} and \ref{SourceFig2}.  

The second example, G323.464$-$00.079, is a hyper-compact \hii region that violates the condition of being optically thin in the continuum \citep{Murphy10}.  Unlike most of our detections, including G314.219+00.344
shown on Figure \ref{SourceFig1}, the continuum flux of G323.464$-$00.079 increases rapidly with increasing frequency across the entire range of these observations.  
In this case the electron temperature derived from the line-to-continuum ratios
is an overestimate.  
Further study of this source using maps of the RRL emission may reveal whether the two
Gaussian components correspond to separate regions of ionized gas.

Accurate values of the continuum and spectral line flux densities will require mapping
and cleaning of the interferometer data.  The full SHRDS survey achieves much better
coverage of the \emph{uv} plane, as it uses two different array configurations and longer
integrations at more and different hour angles.  Thus the values on columns 5 and 7 of 
Table \ref{Table:ResultsTable} will be revised when better data become available.
But the line to continuum ratios, and hence the electron temperatures, can be determined
quickly from the \emph{uv} data alone, as shown in the lower right panels of Figures
\ref{SourceFig1} and \ref{SourceFig2}.

\section{Summary and Conclusions}

The SHRDS Pilot Study discovered 36 \HII regions with radio recombination lines, a detection rate of about 66\%. All known \HII regions included in the observing schedule were also successfully re-detected. More than one third (15 out of 35) of the newly discovered \HII regions are located in the outer Galaxy where existing catalogs of \HII regions are not very complete.  For the sources with RRL detections in the literature, the line strengths, center velocities, and line widths are in good agreement with published values.  The rms of the differences in centre velocities divided by their errors is 1.1, and the 
rms of the differences in FWHM divided by their errors is 1.4.  The only 
velocity parameter that differs from its corresponding value in the literature by more than two sigma
is the FWHM of the line in G295.748$-$00.207, which we find as 23.9$\pm$3.5 \kms
vs. \citet{CH87} value of 35 \kms. \citet{CH87} do not give an error in the
FWHM, but their channel spacing is 2.3 \kms.  

{\bf
  A surprising result of these pilot observations is the low detection rate for candidate \HII regions in the
third quadrant compared with those in the fourth quadrant of the Galaxy.  This may be caused in part by a selection
bias against sources with large angular sizes.  The interferometer is not sensitive to brightness that is smoothly
spread over angular sizes larger than about 6\arcmin, as noted in section 3, paragraph 1.  Thus nearby 
regions with large radii will show lower flux densities for the interferometer than they would to a single dish
telescope.
}

   The main goal of these pilot experiments was to demonstrate the efficiency and sensitivity of the ATCA for detecting RRL emission from Galactic \hii regions using the CABB system.  The secondary objective was to determine the best placement for the CABB zoom bands, given the
varying system temperature and interference environment of the ATCA, and the typical emission spectra of a sample of \hii region candidates.  
The result of the two epochs of observations indicates that the SHRDS should concentrate on the higher frequencies available with the 4 cm (C/X-band) receiver.  These are generally easier to calibrate, more sensitive to RRLs from typical \hii region candidates, and the resolution is better at shorter
wavelengths.

The observing strategy for the SHRDS was also a subject of experimentation in the pilot project.  For candidates near the detection threshold, an efficient strategy is to make many short observations, at widely spaced hour angles so as to get the best
coverage of the $uv$ plane.  These three or four minute integrations are not long enough to detect RRLs, but with five to ten such ``snapshots'' a continuum map with fair to good dynamic range can be made.  Based on the strength of the source(s) found
on the continuum map, we can estimate accurately the expected RRL 
line strength.  The total telescope time available can then be apportioned among the candidates so as to optimise the line detection rate.

\acknowledgements
This research has made use of of NASA's Astrophysics Data System; the SIMBAD database and VizieR catalogue access tool, CDS, Strasbourg, France; and matplotlib for python \citep{matplotlib}.

The National Radio Astronomy Observatory is a facility of the National Science Foundation operated under cooperative agreement by Associated Universities, Inc.
The Australia Telescope Compact Array is part of the Australia Telescope National Facility which is funded by the Australian Government for operation as a National Facility managed by CSIRO.

\startlongtable
\begin{deluxetable}{lcllccllc}
\tablecolumns{9}
\tablewidth{0pt}
\tablecaption{Detected RRL Gaussian Parameters \label{Table:ResultsTable} }

\tablehead{
\colhead{Source}        &       \colhead{\Hna}& \colhead{VLSR}&      \colhead{FWHM}&      \colhead{S$_c$}&   \colhead{RMS}& \colhead{S$_L$}& \colhead{T$_e$}& \colhead{SNR} \\
\colhead{Name}  &       \colhead{(n)}&  \colhead{(\kms)}&       \colhead{(\kms)}&       \colhead{(mJy)}&        \colhead{(mJy)}& \colhead{(mJy)}& \colhead{(K)}& \colhead{$\sigma$}   }

\startdata
G$213.833+00.618$         &      S        &       $53.2   \pm    20.2    $&      33.5     $\pm$  20.9     &      N/A      &      1.4      &      1.9      $\pm$  2.3      &      N/A                      &      3.5             \\
G$230.354-00.597$         &      S        &       $69.0   \pm    16.1    $&      35.0     $\pm$  16.7     &      N/A      &      1.5      &      2.1      $\pm$  2.0      &      N/A                      &      3.7             \\
G$290.012-00.867$         &      S        &       $14.7   \pm    8.3     $&      24.9     $\pm$  8.6      &      N/A      &      1.8      &      3.6      $\pm$  2.5      &      N/A                      &      4.5             \\
G$290.323-02.984$         &      S        &       $-17.7   \pm    8.0     $&      28.3     $\pm$  8.3      &      N/A      &      1.5      &      3.5      $\pm$  2.1      &      N/A                      &      5.6             \\
G$290.385-01.042$         &      S        &       $9.91   \pm    4.6     $&      13.2     $\pm$  4.7      &      N/A      &      1.9      &      4.0      $\pm$  2.9      &      N/A                      &      3.4             \\
G$290.674-00.133$         &      S        &       $19.3   \pm    5.2     $&      23.2     $\pm$  5.3      &      N/A      &      1.4      &      5.2      $\pm$  2.4      &      N/A                      &      8.0             \\
G$291.596-00.239$         &      S        &       $11.4   \pm    5.1     $&      44.2     $\pm$  5.4      &      N/A      &      2.5      &      15.3     $\pm$  3.7      &      8300     $\pm$  1200     &      18.             \\
G$292.889-00.831$         &      S        &       $21.8   \pm    7.1     $&      30.0     $\pm$  7.3      &      N/A      &      1.4      &      4.8      $\pm$  2.3      &      N/A                      &      8.4             \\
G$293.936-00.873$         &      S        &       $36.6   \pm    1.4     $&      22.3     $\pm$  1.5      &      N/A      &      2.0      &      16.4     $\pm$  2.2      &      9400     $\pm$  400      &      17             \\
G$293.994-00.934$         &      S        &       $46.5   \pm    1.7     $&      25.6     $\pm$  1.8      &      N/A      &      1.3      &      18.6     $\pm$  2.6      &      9600     $\pm$  400      &      32            \\
G$294.988-00.538$         &      S        &       $39.8   \pm    1.9     $&      27.4     $\pm$  2.0      &      N/A      &      1.2      &      9.7      $\pm$  1.4      &      10000    $\pm$  400      &      19             \\
G$295.275-00.255$         &      S        &       $30.1   \pm    6.2     $&      30.6     $\pm$  6.3      &      N/A      &      1.6      &      5.6      $\pm$  2.3      &      7900     $\pm$  4500     &      8.7             \\
G$295.748-00.207$         &      S        &       $23.3   \pm    3.5     $&      23.9     $\pm$  3.5      &      N/A      &      1.6      &      7.2      $\pm$  2.1      &      7300     $\pm$  1800     &      9.8             \\
G$297.248-00.754$         &      S        &       $22.6   \pm    1.5     $&      24.4     $\pm$  1.6      &      N/A      &      1.9      &      22.6     $\pm$  2.9      &      7200     $\pm$  400      &      26            \\
         &      107      &      $24.4    \pm    2.7     $&      23.1     $\pm$  2.8      &      369      &      7.2      &      22.0     $\pm$  5.0      &      7700     $\pm$  3200     &      3.1             \\
         &      106      &      $22.0    \pm    2.8     $&      27.6     $\pm$  2.8      &      365      &      7.0      &      21.0     $\pm$  4.0      &      6900     $\pm$  2300     &      3.0             \\
         &      105      &      $22.3    \pm    2.6     $&      25.8     $\pm$  2.7      &      359      &      6.9      &      20.0     $\pm$  4.0      &      7700     $\pm$  2700     &      3.0             \\
         &      104      &      $20.4    \pm    2.3     $&      26.0     $\pm$  2.3      &      356      &      8.3      &      25.0     $\pm$  4.0      &      6500     $\pm$  1900     &      3.1             \\
         &      103      &      $22.3    \pm    2.8     $&      25.3     $\pm$  2.9      &      350      &      7.0      &      21.0     $\pm$  4.0      &      7700     $\pm$  3000     &      3.1             \\
G$297.626-00.906$         &      S        &       $31.3   \pm    2.1     $&      28.4     $\pm$  2.2      &      N/A      &      1.5      &      16.1     $\pm$  2.5      &      7900     $\pm$  400      &      26            \\
G$298.473+00.104$         &      S        &       $32.9   \pm    2.0     $&      23.5     $\pm$  2.1      &      N/A      &      2.0      &      12.2     $\pm$  2.2      &      7900     $\pm$  400      &      13             \\
G$298.669+00.064$         &      S        &       $24.1   \pm    2.4     $&      14.1     $\pm$  2.4      &      N/A      &      2.2      &      11.4     $\pm$  3.9      &      11000    $\pm$  3800     &      8.7             \\
G$300.972+00.994$         &      S        &       $-34.5  \pm    7.2     $&      39.1     $\pm$  7.5      &      N/A      &      1.7      &      3.8      $\pm$  1.4      &      5600     $\pm$  2600     &      6.3             \\
G$300.983+01.117$         &      S        &       $-42.0  \pm    0.4     $&      26.5     $\pm$  0.4      &      N/A      &      3.3      &      83.8     $\pm$  2.8      &      6000     $\pm$  880      &      26.0            \\
         &      107      &      $-42.4   \pm    0.8     $&      25.6     $\pm$  0.9      &      1360     &      12.3     &      86.0     $\pm$  5.0      &      6700     $\pm$  650      &      7.0             \\
         &      106      &      $-43.0   \pm    0.9     $&      25.2     $\pm$  0.9      &      1330     &      12.6     &      86.0     $\pm$  6.0      &      6800     $\pm$  690      &      6.8             \\
         &      105      &      $-42.7   \pm    0.8     $&      27.2     $\pm$  0.8      &      1290     &      13.2     &      91.0     $\pm$  5.0      &      6100     $\pm$  530      &      6.9             \\
         &      104      &      $-41.9   \pm    1.0     $&      26.2     $\pm$  1.0      &      1260     &      13.3     &      83.0     $\pm$  6.0      &      6900     $\pm$  780      &      6.3             \\
         &      103      &      $-42.0   \pm    1.0     $&      25.7     $\pm$  1.0      &      1220     &      12.6     &      85.0     $\pm$  6.0      &      6800     $\pm$  770      &      6.8             \\
         &      96       &      $-42.2   \pm    1.0     $&      26.0     $\pm$  1.0      &      879      &      22.5     &      81.0     $\pm$  6.0      &      6500     $\pm$  720      &      3.6             \\
         &      94       &      $-41.2   \pm    1.0     $&      24.9     $\pm$  1.0      &      809      &      20.0     &      75.0     $\pm$  6.0      &      7100     $\pm$  850      &      3.8             \\
         &      93       &      $-40.6   \pm    1.1     $&      26.6     $\pm$  1.1      &      750      &      17.7     &      73.0     $\pm$  6.0      &      6600     $\pm$  810      &      4.2             \\
G$313.671-00.104$         &      S        &       $-54.6  \pm    1.4     $&      24.6     $\pm$  1.5      &      N/A      &      1.6      &      12.0     $\pm$  1.4      &      6700     $\pm$  400      &      17             \\
G$313.790+00.706$         &      S        &       $-57.2  \pm    0.9     $&      22.6     $\pm$  0.9      &      N/A      &      2.6      &      38.7     $\pm$  3.0      &      7000     $\pm$  450      &      32            \\
         &      101      &      $-57.6   \pm    1.9     $&      20.2     $\pm$  1.9      &      436      &      11.1     &      41.0     $\pm$  8.0      &      6800     $\pm$  2200     &      3.7             \\
         &      97       &      $-56.7   \pm    1.8     $&      20.6     $\pm$  1.8      &      399      &      11.2     &      38.0     $\pm$  6.0      &      7300     $\pm$  2100     &      3.5             \\
         &      96       &      $-56.8   \pm    2.6     $&      26.9     $\pm$  2.6      &      396      &      11.8     &      35.0     $\pm$  6.0      &      6400     $\pm$  2100     &      3.0             \\
         &      95       &      $-55.0   \pm    2.2     $&      26.4     $\pm$  2.3      &      386      &      11.7     &      35.0     $\pm$  6.0      &      6600     $\pm$  1800     &      3.1             \\
         &      94       &      $-58.5   \pm    1.6     $&      20.7     $\pm$  1.6      &      372      &      12.8     &      41.0     $\pm$  6.0      &      7100     $\pm$  1700     &      3.2             \\
         &      89       &      $-55.7   \pm    1.7     $&      21.7     $\pm$  1.8      &      330      &      11.9     &      39.0     $\pm$  6.0      &      7400     $\pm$  1900     &      3.3             \\
         &      87       &      $-57.0   \pm    1.8     $&      21.3     $\pm$  1.9      &      306      &      11.9     &      40.0     $\pm$  7.0      &      7500     $\pm$  2200     &      3.4             \\
         &      85       &      $-56.2   \pm    1.7     $&      22.2     $\pm$  1.8      &      294      &      13.8     &      42.0     $\pm$  6.0      &      7100     $\pm$  1800     &      3.1             \\
G$314.219+00.344$         &      S        &       $-62.5  \pm    0.4     $&      20.0     $\pm$  0.5      &      N/A      &      3.5      &      66.4     $\pm$  3.1      &      6500     $\pm$  640      &      38            \\
         &      101      &      $-63.0   \pm    1.0     $&      21.0     $\pm$  1.0      &      706      &      12.8     &      81.0     $\pm$  7.0      &      5600     $\pm$  780      &      6.3             \\
         &      97       &      $-62.7   \pm    0.9     $&      18.1     $\pm$  0.9      &      534      &      12.3     &      71.0     $\pm$  7.0      &      6200     $\pm$  940      &      5.8             \\
         &      96       &      $-61.7   \pm    1.4     $&      20.6     $\pm$  1.4      &      495      &      12.2     &      55.0     $\pm$  7.0      &      6700     $\pm$  1400     &      4.6             \\
         &      95       &      $-63.1   \pm    1.2     $&      18.4     $\pm$  1.2      &      468      &      12.1     &      63.0     $\pm$  8.0      &      6500     $\pm$  1300     &      5.2             \\
         &      94       &      $-62.1   \pm    1.2     $&      21.8     $\pm$  1.2      &      438      &      12.9     &      59.0     $\pm$  6.0      &      5700     $\pm$  990      &      4.6             \\
         &      89       &      $-61.6   \pm    1.2     $&      21.2     $\pm$  1.2      &      401      &      12.5     &      57.0     $\pm$  6.0      &      6500     $\pm$  1100     &      4.6             \\
         &      87       &      $-62.7   \pm    1.0     $&      18.6     $\pm$  1.0      &      389      &      13.0     &      67.0     $\pm$  7.0      &      6700     $\pm$  1100     &      5.1             \\
         &      85       &      $-62.5   \pm    1.5     $&      20.1     $\pm$  1.6      &      360      &      15.3     &      65.0     $\pm$  10.0     &      6400     $\pm$  1600     &      4.3             \\
G$315.312-00.273$         &      S        &       $14.2   \pm    1.4     $&      24.0     $\pm$  1.4      &      N/A      &      3.6      &      30.0     $\pm$  3.6      &      7700     $\pm$  400      &      18             \\
         &      101      &      $10.7    \pm    2.1     $&      18.1     $\pm$  2.2      &      414      &      11.5     &      37.0     $\pm$  9.0      &      7800     $\pm$  3300     &      3.3             \\
         &      97       &      $14.1    \pm    2.4     $&      22.4     $\pm$  2.4      &      377      &      11.9     &      36.0     $\pm$  7.0      &      6900     $\pm$  2500     &      3.1             \\
         &      85       &      $15.2    \pm    4.5     $&      33.4     $\pm$  4.7      &      240      &      14.9     &      22.0     $\pm$  6.0      &      7400     $\pm$  3900     &      1.5             \\
G$316.516-00.600$         &      S        &       $-45.6  \pm    0.9     $&      19.9     $\pm$  0.9      &      N/A      &      1.7      &      19.0     $\pm$  1.7      &      5900     $\pm$  400      &      22            \\
         &      97       &      $-46.2   \pm    2.0     $&      18.1     $\pm$  2.0      &      142      &      6.1      &      20.0     $\pm$  4.0      &      5900     $\pm$  2300     &      3.3             \\
         &      96       &      $-45.4   \pm    2.0     $&      19.4     $\pm$  2.0      &      140      &      6.5      &      20.0     $\pm$  4.0      &      5500     $\pm$  2000     &      3.2             \\
         &      94       &      $-46.2   \pm    1.8     $&      18.2     $\pm$  1.9      &      135      &      6.6      &      21.0     $\pm$  4.0      &      5900     $\pm$  2100     &      3.2             \\
         &      89       &      $-42.7   \pm    1.8     $&      19.8     $\pm$  1.8      &      115      &      6.7      &      21.0     $\pm$  3.0      &      5600     $\pm$  1700     &      3.2             \\
         &      87       &      $-45.8   \pm    1.9     $&      20.3     $\pm$  1.9      &      106      &      6.9      &      21.0     $\pm$  4.0      &      5300     $\pm$  1700     &      3.2             \\
G$317.861+00.160$         &      S        &       $1.53   \pm    0.9     $&      22.9     $\pm$  0.9      &      N/A      &      2.7      &      55.6     $\pm$  4.6      &      7600     $\pm$  430      &      44            \\
         &      101      &      $1.68    \pm    1.7     $&      24.4     $\pm$  1.7      &      713      &      13.0     &      55.0     $\pm$  7.0      &      6900     $\pm$  1500     &      4.2             \\
         &      97       &      $2.22    \pm    1.6     $&      25.0     $\pm$  1.6      &      648      &      13.2     &      55.0     $\pm$  7.0      &      7000     $\pm$  1400     &      4.2             \\
         &      96       &      $1.49    \pm    1.3     $&      19.6     $\pm$  1.4      &      631      &      13.8     &      66.0     $\pm$  9.0      &      7400     $\pm$  1600     &      4.8             \\
         &      95       &      $1.24    \pm    1.5     $&      20.7     $\pm$  1.5      &      614      &      13.3     &      54.0     $\pm$  8.0      &      8400     $\pm$  2000     &      4.1             \\
         &      94       &      $2.88    \pm    1.9     $&      28.5     $\pm$  2.0      &      597      &      14.4     &      53.0     $\pm$  7.0      &      6600     $\pm$  1400     &      3.7             \\
         &      89       &      $0.942   \pm    1.4     $&      21.3     $\pm$  1.5      &      506      &      13.9     &      55.0     $\pm$  7.0      &      8200     $\pm$  1700     &      4.0             \\
         &      87       &      $0.801   \pm    1.7     $&      23.8     $\pm$  1.8      &      466      &      14.8     &      52.0     $\pm$  7.0      &      7800     $\pm$  1800     &      3.5             \\
         &      85       &      $-0.447  \pm    2.4     $&      29.8     $\pm$  2.5      &      431      &      16.5     &      49.0     $\pm$  8.0      &      6800     $\pm$  1900     &      3.0             \\
G$318.248+00.151$         &      S        &       $-39.9  \pm    2.0     $&      19.0     $\pm$  2.0      &      N/A      &      2.8      &      17.0     $\pm$  3.7      &      5900     $\pm$  740      &      12             \\
         &      101      &      $-43.0   \pm    1.7     $&      12.2     $\pm$  1.7      &      175      &      11.0     &      34.0     $\pm$  10.0     &      5500     $\pm$  3000     &      3.2             \\
G$319.229+00.225$         &      S        &       $-66.1  \pm    1.8     $&      20.9     $\pm$  1.8      &      N/A      &      1.9      &      12.9     $\pm$  2.2      &      6800     $\pm$  400      &      14             \\
G$323.449+00.095$         &      S        &       $-75.1  \pm    1.7     $&      22.5     $\pm$  1.7      &      N/A      &      2.0      &      13.8     $\pm$  2.1      &      5900     $\pm$  400      &      15             \\
G$323.464-00.079$         &      S        &       $-68.8  \pm    1.2     $&      38.3     $\pm$  1.2      &      N/A      &      1.3      &      28.0     $\pm$  2.2      &      11000    $\pm$  1700      &      45            \\
         &      S2       &      $-68.1  \pm     1.2     $&      27.7     $\pm$  2       &      N/A      &      1.3       &        24     $\pm$  2.2      &                               &      21            \\
         &      S2       &      $-77.2  \pm     1.2     $&      105.5    $\pm$  10      &      N/A      &      1.3       &       6.5     $\pm$  2.2      &                               &      4.9            \\
         &      101      &      $-66.4   \pm    2.5     $&      31.2     $\pm$  2.6      &      589      &      6.3      &      20.0     $\pm$  3.0      &      11000    $\pm$  2900     &      3.2             \\
         &      97       &      $-67.1   \pm    1.7     $&      27.7     $\pm$  1.8      &      636      &      6.2      &      26.0     $\pm$  3.0      &      12000    $\pm$  2300     &      4.3             \\
         &      96       &      $-67.1   \pm    1.8     $&      27.3     $\pm$  1.9      &      649      &      6.6      &      26.0     $\pm$  3.0      &      13000    $\pm$  2700     &      4.0             \\
         &      95       &      $-67.7   \pm    2.0     $&      27.3     $\pm$  2.1      &      661      &      6.5      &      24.0     $\pm$  3.0      &      14000    $\pm$  3400     &      3.8             \\
         &      94       &      $-69.3   \pm    1.9     $&      30.7     $\pm$  1.9      &      673      &      7.1      &      27.0     $\pm$  3.0      &      12000    $\pm$  2200     &      3.9             \\
         &      89       &      $-68.5   \pm    1.6     $&      35.2     $\pm$  1.6      &      746      &      7.7      &      35.0     $\pm$  3.0      &      11000    $\pm$  1500     &      4.6             \\
         &      87       &      $-68.6   \pm    1.3     $&      36.6     $\pm$  1.3      &      770      &      8.2      &      40.0     $\pm$  2.0      &      11000    $\pm$  1100     &      4.9             \\
         &      85       &      $-68.8   \pm    1.4     $&      36.9     $\pm$  1.4      &      804      &      9.2      &      42.0     $\pm$  3.0      &      11000    $\pm$  1200     &      4.7             \\
G$323.743-00.249$         &      S        &       $-47.3  \pm    0.7     $&      19.2     $\pm$  0.7      &      N/A      &      2.2      &      22.8     $\pm$  1.8      &      6000     $\pm$  400      &      20            \\
         &      101      &      $-47.3   \pm    1.9     $&      17.0     $\pm$  1.9      &      165      &      6.9      &      22.0     $\pm$  5.0      &      5800     $\pm$  2300     &      3.2             \\
         &      97       &      $-46.7   \pm    1.6     $&      17.7     $\pm$  1.6      &      160      &      6.8      &      25.0     $\pm$  4.0      &      5500     $\pm$  1700     &      3.8             \\
         &      95       &      $-47.7   \pm    2.0     $&      19.2     $\pm$  2.1      &      157      &      6.8      &      21.0     $\pm$  4.0      &      6100     $\pm$  2300     &      3.2             \\
         &      89       &      $-47.8   \pm    1.4     $&      17.3     $\pm$  1.4      &      147      &      7.1      &      27.0     $\pm$  4.0      &      6300     $\pm$  1600     &      3.8             \\
         &      87       &      $-47.6   \pm    1.6     $&      21.7     $\pm$  1.7      &      141      &      7.3      &      26.0     $\pm$  4.0      &      5400     $\pm$  1400     &      3.6             \\
         &      85       &      $-46.1   \pm    2.3     $&      19.4     $\pm$  2.4      &      138      &      7.6      &      25.0     $\pm$  6.0      &      6600     $\pm$  3000     &      3.3             \\
G$323.806+00.021$         &      S        &       $-58.9  \pm    0.9     $&      21.6     $\pm$  0.9      &      N/A      &      2.7      &      31.6     $\pm$  2.7      &      7200     $\pm$  400      &      24            \\
         &      101      &      $-57.4   \pm    2.0     $&      18.8     $\pm$  2.0      &      373      &      10.0     &      35.0     $\pm$  7.0      &      7300     $\pm$  2700     &      3.5             \\
         &      97       &      $-57.0   \pm    1.7     $&      18.4     $\pm$  1.7      &      333      &      10.0     &      36.0     $\pm$  6.0      &      7400     $\pm$  2300     &      3.6             \\
         &      96       &      $-59.3   \pm    2.1     $&      25.4     $\pm$  2.1      &      320      &      10.4     &      37.0     $\pm$  6.0      &      5400     $\pm$  1400     &      3.6             \\
         &      89       &      $-59.4   \pm    1.8     $&      21.0     $\pm$  1.9      &      248      &      11.0     &      35.0     $\pm$  6.0      &      6700     $\pm$  2000     &      3.2             \\
G$323.936-00.037$         &      S        &       $-57.3  \pm    4.5     $&      24.5     $\pm$  4.6      &      N/A      &      2.4      &      13.8     $\pm$  5.2      &      5600     $\pm$  2600     &      12.7             \\
G$324.662-00.331$         &      S        &       $-47.8  \pm    1.7     $&      21.3     $\pm$  1.8      &      N/A      &      2.9      &      28.7     $\pm$  4.8      &      7200     $\pm$  400      &      20             \\
G$325.108+00.054$         &      S        &       $-67.8  \pm    2.7     $&      20.4     $\pm$  2.7      &      N/A      &      3.2      &      11.5     $\pm$  3.1      &      6700     $\pm$  1600     &      7.3             \\
G$325.354-00.035$         &      S        &       $-63.8  \pm    0.6     $&      $34.4     \pm  1.5$      &      N/A      &      1.4      &      $15     \pm  1.5$      &      $6800     \pm  800$      &      26             \\
 & 102 & -63.3 $\pm$ 1.4 & 25.2 $\pm$ 3.4 & 251 & 4.4 & 17 $\pm$ 2.0 & 7232 $\pm$ 970 & 8.7 \\
 & 101 & -68.8 $\pm$ 2.0 & 31.3 $\pm$ 4.7 & 252 & 4.6 & 14 $\pm$ 1.9 & 7168 $\pm$ 1100 & 7.7 \\
 & 100 & -62.7 $\pm$ 2.0 & 36.8 $\pm$ 4.8 & 246 & 5.2 & 17 $\pm$ 2.0 & 5335 $\pm$ 690 & 8.9 \\
 &  98 & -59.1 $\pm$ 1.1 & 19.6 $\pm$ 2.7 & 235 & 6.1 & 26 $\pm$ 3.1 & 6630 $\pm$ 920 & 8.4 \\
 &  97 & -61.4 $\pm$ 2.4 & 37.9 $\pm$ 6.0 & 228 & 5.5 & 19 $\pm$ 2.1 & 4934 $\pm$ 800 & 9.2 \\
 &  96 & -60.2 $\pm$ 1.3 & 19.4 $\pm$ 3.4 & 222 & 6.0 & 24 $\pm$ 3.2 & 7142 $\pm$ 1300 & 7.7 \\
 &  95 & -62.5 $\pm$ 2.0 & 27.4 $\pm$ 4.7 & 217 & 5.6 & 16 $\pm$ 2.4 & 7574 $\pm$ 1400 & 6.7 \\
 &  94 & -65.4 $\pm$ 1.8 & 32.2 $\pm$ 4.3 & 211 & 5.0 & 17 $\pm$ 2.0 & 6244 $\pm$ 830 & 8.7 \\
 &  89 & -63.8 $\pm$ 2.1 & 31.3 $\pm$ 5.1 & 183 & 4.6 & 13 $\pm$ 1.8 & 8501 $\pm$ 1400 & 7.1 \\
 &  88 & -60.6 $\pm$ 1.6 & 28.9 $\pm$ 3.7 & 178 & 4.6 & 17 $\pm$ 1.9 & 7130 $\pm$ 910 & 9.0 \\
 &  87 & -60.9 $\pm$ 1.3 & 30.1 $\pm$ 3.1 & 172 & 4.6 & 22 $\pm$ 1.9 & 5746 $\pm$ 570 & 11.4 \\
 &  86 & -56.5 $\pm$ 1.8 & 25.9 $\pm$ 4.3 & 167 & 5.7 & 18 $\pm$ 2.5 & 7869 $\pm$ 1350 & 7.0 \\
 &  85 & -59.8 $\pm$ 1.3 & 21.7 $\pm$ 3.1 & 162 & 5.2 & 21 $\pm$ 2.5 & 8035 $\pm$ 1140 & 8.2 \\
G$326.721+00.773$         &      S        &       $-40.6  \pm    0.9     $&      22.6     $\pm$  0.9      &      N/A      &      4.9      &      45.5     $\pm$  3.7      &      6300     $\pm$  400      &      20             \\
         &      101      &      $-40.9   \pm    2.0     $&      18.8     $\pm$  2.0      &      458      &      14.3     &      51.0     $\pm$  11.0     &      6300     $\pm$  2300     &      3.6             \\
         &      97       &      $-40.6   \pm    2.0     $&      23.8     $\pm$  2.0      &      470      &      15.4     &      53.0     $\pm$  9.0      &      5700     $\pm$  1500     &      3.5             \\
         &      96       &      $-40.2   \pm    2.0     $&      25.4     $\pm$  2.1      &      480      &      16.4     &      56.0     $\pm$  9.0      &      5400     $\pm$  1400     &      3.5             \\
         &      95       &      $-40.3   \pm    2.2     $&      22.0     $\pm$  2.2      &      466      &      15.7     &      46.0     $\pm$  9.0      &      7200     $\pm$  2400     &      3.0             \\
G$326.890-00.277$         &      S        &       $-44.2  \pm    0.7     $&      19.1     $\pm$  0.7      &      N/A      &      3.5      &      63.2     $\pm$  4.5      &      6200     $\pm$  440      &      35            \\
         &      101      &      $-45.4   \pm    1.8     $&      19.4     $\pm$  1.8      &      487      &      13.0     &      49.0     $\pm$  9.0      &      6700     $\pm$  2000     &      3.8             \\
         &      97       &      $-43.7   \pm    1.2     $&      18.1     $\pm$  1.2      &      465      &      13.4     &      65.0     $\pm$  8.0      &      6000     $\pm$  1200     &      4.9             \\
         &      96       &      $-45.0   \pm    1.8     $&      22.6     $\pm$  1.9      &      456      &      13.8     &      53.0     $\pm$  8.0      &      5900     $\pm$  1600     &      3.9             \\
         &      95       &      $-45.3   \pm    1.6     $&      22.2     $\pm$  1.7      &      450      &      13.6     &      54.0     $\pm$  8.0      &      6000     $\pm$  1400     &      4.0             \\
         &      94       &      $-43.4   \pm    1.2     $&      18.3     $\pm$  1.2      &      445      &      15.7     &      62.0     $\pm$  8.0      &      6500     $\pm$  1300     &      4.0             \\
         &      89       &      $-43.7   \pm    0.9     $&      18.7     $\pm$  0.9      &      415      &      13.9     &      75.0     $\pm$  7.0      &      6000     $\pm$  880      &      5.4             \\
         &      87       &      $-44.3   \pm    0.9     $&      20.5     $\pm$  1.0      &      408      &      14.3     &      77.0     $\pm$  7.0      &      5600     $\pm$  800      &      5.5             \\
         &      85       &      $-43.7   \pm    0.8     $&      17.9     $\pm$  0.8      &      396      &      15.1     &      80.0     $\pm$  7.0      &      6400     $\pm$  900      &      5.3             \\
G$326.916-01.100$         &      S        &       $-49.8  \pm    1.8     $&      21.0     $\pm$  1.9      &      N/A      &      4.4      &      18.9     $\pm$  3.4      &      6800     $\pm$  450      &      8.8             \\
G$327.313-00.536$         &      S        &       $-48.5  \pm    0.2     $&      28.1     $\pm$  0.3      &      N/A      &      33.1     &      2280.0   $\pm$  41.7     &      6100     $\pm$  890      &      163            \\
         &      102      &      $-48.8   \pm    0.4     $&      28.8     $\pm$  0.4      &      26600    &      200.0    &      2020.0   $\pm$  56.0     &      5900     $\pm$  400      &      10.0            \\
         &      101      &      $-48.9   \pm    0.3     $&      28.5     $\pm$  0.3      &      27000    &      211.0    &      2128.0   $\pm$  47.0     &      5900     $\pm$  400      &      10.0            \\
         &      97       &      $-48.7   \pm    0.3     $&      28.0     $\pm$  0.3      &      25100    &      239.0    &      2226.0   $\pm$  46.0     &      6100     $\pm$  400      &      9.3             \\
         &      96       &      $-48.7   \pm    0.3     $&      28.0     $\pm$  0.3      &      24600    &      238.0    &      2293.0   $\pm$  48.0     &      6000     $\pm$  400      &      9.6             \\
         &      95       &      $-48.6   \pm    0.3     $&      28.2     $\pm$  0.3      &      24100    &      234.0    &      2289.0   $\pm$  50.0     &      6100     $\pm$  400      &      9.8             \\
         &      94       &      $-48.5   \pm    0.3     $&      27.6     $\pm$  0.3      &      23500    &      303.0    &      2315.0   $\pm$  53.0     &      6200     $\pm$  400      &      7.6             \\
         &      89       &      $-48.6   \pm    0.3     $&      28.1     $\pm$  0.3      &      20300    &      270.0    &      2420.0   $\pm$  50.0     &      6000     $\pm$  400      &      9.0             \\
         &      87       &      $-48.3   \pm    0.3     $&      27.6     $\pm$  0.3      &      19100    &      282.0    &      2479.0   $\pm$  53.0     &      6100     $\pm$  400      &      8.8             \\
         &      85       &      $-47.9   \pm    0.3     $&      27.3     $\pm$  0.3      &      18000    &      290.0    &      2471.0   $\pm$  53.0     &      6200     $\pm$  400      &      8.5             \\
G$327.401+00.484$         &      S        &       $-76.3  \pm    1.7     $&      17.7     $\pm$  1.7      &      N/A      &      2.6      &      31.5     $\pm$  6.1      &      6000     $\pm$  570      &      23            \\
         &      97       &      $-77.1   \pm    1.6     $&      17.0     $\pm$  1.7      &      229      &      13.7     &      47.0     $\pm$  9.0      &      4500     $\pm$  1500     &      3.5             \\
G$327.555-00.829$         &      S        &       $-41.7  \pm    2.6     $&      24.8     $\pm$  2.6      &      N/A      &      4.0      &      16.5     $\pm$  3.5      &      7500     $\pm$  950      &      9.2             \\
G$327.714+00.577$         &      S        &       $-47.4  \pm    2.4     $&      21.9     $\pm$  2.5      &      N/A      &      2.1      &      12.0     $\pm$  2.7      &      6600     $\pm$  1000     &      12             \\
G$327.763+00.163$         &      S        &       $-92.8  \pm    1.2     $&      21.6     $\pm$  1.2      &      N/A      &      2.6      &      30.1     $\pm$  3.4      &      6800     $\pm$  400       &      24            \\
         &      97       &      $-93.5   \pm    2.1     $&      19.8     $\pm$  2.1      &      297      &      10.3     &      33.0     $\pm$  7.0      &      6700     $\pm$  2500     &      3.3             \\
         &      94       &      $-93.5   \pm    1.8     $&      21.2     $\pm$  1.8      &      265      &      10.7     &      37.0     $\pm$  6.0      &      5700     $\pm$  1600     &      3.5             \\
         &      87       &      $-92.9   \pm    2.0     $&      19.2     $\pm$  2.0      &      191      &      10.8     &      35.0     $\pm$  7.0      &      6100     $\pm$  2300     &      3.3             \\
\enddata
\end{deluxetable}

\end{document}